\documentclass[final,3p,times]{elsarticle}

\usepackage{amssymb}
\usepackage{amsthm}
\usepackage{amsmath}
\usepackage[normalem]{ulem}

\usepackage{lineno}
\usepackage{multirow}
\usepackage[toc,page,title,titletoc,header]{appendix} 
\usepackage{graphicx}
\usepackage{booktabs}
\usepackage{hhline}
\usepackage{threeparttable}
\usepackage{caption}
\usepackage{color}
\usepackage{subcaption}
\usepackage{hyperref}
\usepackage{setspace}
\DeclareMathAlphabet{\pazocal}{OMS}{zplm}{m}{n}
\journal{Arxiv}

\begin{document}
\begin{frontmatter}



\title{LSHR-Net: a hardware-friendly solution for high-resolution computational imaging using a mixed-weights neural network}


\author[label1]{Fangliang Bai}
\author[label2]{Jinchao Liu}
\author[label4,label5]{Xiaojuan Liu}
\author[label3]{Margarita Osadchy}
\author[label4]{Chao~Wang}
\author[label1]{Stuart J. Gibson}
\address[label1]{School of Physical Sciences, University of Kent, Canterbury, Kent, UK, CT2 7NH}
\address[label2]{College of Artificial Intelligence, Nankai University, Tianjin, China, 300071}
\address[label3]{Department of Computer Science, University of Haifa, Mount Carmel, Haifa, Israel}
\address[label4]{School of Engineering and Digital Arts, University of Kent, Canterbury Kent, UK, CT2 7NT}
\address[label5]{School of Physics and Optoelectronics Engineering, Shandong University of Technology, Zibo, China, 255049}

\begin{abstract}
Recent work showed neural-network based approaches to reconstructing images from compressively sensed measurements offer significant improvements in accuracy and signal compression. Such methods can dramatically boost the capability of computational imaging hardware.
However, to date, there have been two major drawbacks:
(1) the high-precision real-valued sensing patterns proposed in the majority of existing works can prove problematic when used with computational imaging hardware such as a digital micromirror sampling device and
(2) the network structures for image reconstruction involve intensive computation, which is also not suitable for hardware deployment.
To address these problems, we propose a novel hardware-friendly solution based on mixed-weights neural networks for computational imaging. In particular, learned binary-weight sensing patterns are tailored to the sampling device.
Moreover, we proposed a recursive network structure for low-resolution image sampling and high-resolution reconstruction scheme. It reduces both the required number of measurements and reconstruction computation by operating convolution on small intermediate feature maps. The recursive structure further reduced the model size, making the network more computationally efficient when deployed with the hardware.
Our method has been validated on benchmark datasets and achieved state of the art reconstruction accuracy. We tested our proposed network in conjunction with a proof-of-concept hardware setup.
\end{abstract}


\end{frontmatter}

\section{Introduction}\label{S:1}
In the context of structural signal recovery, the task of image reconstruction from the compressive sampling has been closely associated with computational  imaging \cite{shapiro2008computational} using a single pixel camera \cite{wakin2006architecture,sank2015video}. Single pixel camera architectures are of particular interest when imaging outside the visible range of the electromagnetic spectrum in cases where detector technology is expensive or difficult to manufacture. This approach to image acquisition involves illuminating an object scene using a sampling device which produces structured light in the form of 2D pseudo-random patterns. For each pattern, the intensity of the back scattered light is measured by a single pixel photo-detector.
In the computational imaging paradigm \cite{wakin2006architecture}, each measurement corresponds to the inner product between a sensing pattern and the image to be reconstructed. This can be formulated as:
\begin{equation}
  y = \Phi x + e
\end{equation}
where $x \in \mathbb{R}^{n}$ is the image rearranged as a vector, $\Phi \in \mathbb{R}^{m \times n}$, \space $m \ll n $, are $m$ random sensing patterns (also concatenated into vector form), $e \in \mathbb{R}^{m}$ are measurement errors and $y \in \mathbb{R}^{m} $ are the measurements. The number of sensing patterns $m$ can be much fewer than the total number of pixels $n$ comprising the reconstructed image, resulting in a measurement ratio of $R=\frac{m}{n}$.

A digital micro-mirror device (DMD) is widely used as the sampling component in single pixel camera architectures and for coded aperture imaging \cite{sun2016single, sun2019single, Lochocki:16, Zhang:17, Sun:16, chiranjan2016implementation}. It contains a 2d array of micro-mirrors (hence the name) and each micro-mirror can be positioned at one of two angles to be in either an activated or inactivated state.
When the array is illuminated by a uniform light source, shifting the micromirrors between states produces different binary sensing patterns, such as random Bernoulli, Hadamard, which are projected onto the object scene of interest.
Given an incident, uniform, light source, shifting mirrors between states produces different binary sensing patterns, such as random Bernoulli, Hadamard, which are used to illuminate the object scene of interest.

To reconstruct signals/images from compressively sampled measurements, Compressed sensing (CS)\cite{donoho2006compressed, candes2006robust}, to be exact sparse optimization methods such as NESTA\cite{Becker2011_NESTA}, ADMM \cite{MAL016} etc. have been proposed and have become the predominant algorithms using in a variety of applications. However, one major drawback of these numerical nonlinear optimization methods is that they often take a few minutes to recover a single large image at good quality.

Deep neural networks (DNNs) have become prevalent in a broad range of image processing tasks \cite{krizhevsky2012imagenet, girshick2014rich, long2015fully, gu2018recent, tu2019survey}. Specifically, DNN has been shown to achieve favorable results in
image recovery \cite{tai2017memnet}. Motivated by this success in image reconstruction tasks, DNNs were subsequently investigated for image reconstruction problems based on compressively sensed image data, \cite{mousavi2015deep, adler2016deep, mousavi2017learning, mousavi2017deepcodec, kulkarni2016reconnet, yao2017dr, xie2017adaptive, xie2017fully, zhao2019visualizing, xie2018full, shi2019image}. These neural network based solutions were reported to outperform the state-of-the-art in compressed sensing algorithms in terms of speed, accuracy and data compression.

Although a variety of different network architectures were proposed, few were deliberately designed to be adaptable to the sensing hardware. To date, there have been two issues that remain to be solved.
First, the real-valued sensing patterns of all existing neural network implementations for this application were stored in 32-bit floating-point format. Although high-precision sensing patterns can be used for software simulation of image sampling on modern GPUs, this is not a realistic representation of sampling using structured light sensing hardware, where instead binary patterns are used to reduce sampling complexity.
Second, previous methods assumed that the sensing patterns and the reconstructed images have the same resolution. Therefore, the size of the recovered image is dependent on the size of the sensing patterns (for dense-connection based methods) or the number of convolutional patch-sampling operations (for convolutional-based methods). For large images, these methods result in large intermediate feature maps and increase the number of operations required for recovering an image. This is because the number of sampling measurements and convolutional computations depends on the size of the feature maps.   
In addition, when the patterns are loaded in hardware, such as a DMD, the maximum reconstruction resolution will be limited by the size of the mirror array (which is fixed) used in the sensing device.

The limitations of previous methods motivated us to design a hardware-friendly deep learning solution, incorporating binary sensing patterns to reconstruct high-resolution images. Previous papers have highlighted the importance of integrating the DNN solutions with hardware \cite{zhao2019visualizing, xie2018full}. In this respect, we go one step further than previous work and provide evidence that our architecture performs well with imaging hardware. 
We propose a new network architecture that:
\begin{enumerate}
    \item Uses a mixed-weights network with sparse binary patterns which lends itself naturally to hardware implementation and can be trained in an end-to-end manner. Unlike floating-point numbers, binary patterns are appropriate for both sampling and measuring hardware. Specifically, the sparse binary patterns can be represented on a DMD without the need for any additional modulation and require less on-board memory usage. Our approach effectively increases the light intensity sensitivity of the single pixel camera (the photo-diode) and the analogue to digital conversion range, compared with methods based on real-valued sensing patterns.
    
    \item Uses a novel sensing-reconstruction scheme, which we term low-resolution sensing with high-resolution reconstruction (LSHR), to directly reconstruct high-resolution images from low-resolution sampled measurements. Given a pattern generated by a DMD of fixed size, the network reconstructs a high-resolution image which has more pixels than the number of micro-mirrors in the array. This low-throughput sampling scheme results in smaller feature maps, and therefore, fewer computational operations are required. Hence, it is more efficient than previously reported methods for use with hardware imaging set-ups.
    
    \item Has a residual-correction sub-net that consists of a chain of recursive residual blocks, where weights are shared between different blocks. Compared with previous methods, our structure further reduces the model size, making it ideal for the limited onboard memory capacity of the hardware (e.g. single pixel camera) while yielding higher reconstruction PSNR accuracy.
    
    \item Achieves state-of-the-art results on benchmark datasets and has been validated on proof-of-concept hardware.
\end{enumerate}
The remainder of this paper is organized as follows: In Section \ref{S:2}, we review the related work on sensing patterns. We describe the design of our proposed network in Section \ref{S:3}. In Section \ref{S:4} we show software simulation results for our model and compare them with existing methods. In Section \ref{S:5}, we present the work of integrating the model with hardware. Finally, in Section \ref{S:6} we conclude our discussion and suggest potential future directions for the work.

\section{Related work} \label{S:2}
The concept of neural network based image reconstruction was first implemented using a fully-connected network \cite{mousavi2015deep}. Thereafter, the problem was approached using convolutional neural networks which avoid the fixed size input image constraint. We organized the related methods \cite{mousavi2015deep,kulkarni2016reconnet,yao2017dr,mousavi2017learning,adler2016deep,mousavi2017deepcodec,xie2017adaptive,xie2017fully,iliadis2018deep,iliadis2016deepbinarymask,zhao2019visualizing,xie2018full,shi2019image} into three categories according to the type of sensing pattern used (randomly generated, learned and binary) and discuss relevant prior work below.

\textit{Networks based on pre-generated (static) patterns}.
 A stacked denoising auto-encoder (SDA) was previously implemented \cite{mousavi2015deep} comprising fully-connected layers. It was trained with measurements acquired by sensing images with pre-generated random Gaussian patterns.
Inspired by SDA, ReconNet \cite{kulkarni2016reconnet} was subsequently proposed. It improved the accuracy by extending the network with additional convolutional layers of different kernel sizes. However the fully-connected layer caused heavy computation and large model size, the sensing area was constrained to small patches of the original image. In the post-processing step, the reconstructed small patches were concatenated to form the whole image. The BM3D \cite{dabov2007image} was then applied to smooth the edges between patches.
The performance of the ReconNet was further improved by DR$^{2}$-Net \cite{yao2017dr}. Here the convolutional layers were replaced with residual blocks which make the network easier to train. But the sensing was still done in small patches.
In contrast to previous methods that used fixed (pre-generated) Gaussian sensing patterns, DeepInverse \cite{mousavi2017learning} used real time generation of random patterns for sampling images.

\textit{Networks based on learned patterns}.
Some of the work described in the previous paragraph has been modified such that the sensing patterns adapt to a particular set of images through a learning process.
The SDA was further adapted to learn the patterns with a fully-connected layer that inputs an image $x$ directly into the network. The fully-connected layer was trained to obtain the measurements $y$ when presented with $x$. This operation can be represented as $y = \sigma(Wx+b)$ where the $\sigma(\cdot)$ is an activation function and $W$ and $b$ are the weights and bias of the fully-connected layer.
A similar structure to SDA was also proposed that employed a fully-connected neural network to implement the block-based compressed sensing \cite{adler2016deep}. The model was trained to jointly optimize the sensing patterns and the network weights.
DeepInverse was also optimized resulting in a new model named DeepCodedec \cite{mousavi2017deepcodec}. It had an encoder-decoder architecture. The network was trained to take measurements from images using several convolutional layers. Unlike SDA, it gradually reduced the dimension of the intermediate feature maps prior to generating the measurements. The efficiency was improved by applying convolutional layers.
The ReconNet was also further improved using learned patterns \cite{xie2017adaptive} and \cite{zhao2019visualizing}. Before training, the fully-connected layer was initialized with random Gaussian patterns. It was then updated during the training. For testing the network, the trained patterns were fixed to perform the sensing. The results showed further improvements in reconstruction accuracy due to learning the patterns. However, the fully-connected layer caused intensive computation and blocking artifacts to appear in the reconstructed images.
To deal with the aforementioned limitations, the authors proposed two networks, \cite{xie2017fully} and \cite{xie2018full}, that sensed images with a convolutional layer with a small stride step to avoid the blocking artifacts.

\textit{Networks based on a binary matrix}.
Neural networks with binary weights were initially designed for image classification tasks, \cite{cour2015binaryconnect, rastegari2016xnor}.
A network for video reconstruction, using binary patterns, was described in \cite{iliadis2018deep}. The network applied a 3D binary sampling matrix to down-sample a sequence of the temporal video frames and learned a non-linear rule, mapping between the measurements and the reconstructed frames via fully-connected layers.
In more recent network, DeepBinaryMask \cite{iliadis2016deepbinarymask},  followed the same strategy of using a binary down-sampling matrix for sensing video frames but introduced a learning procedure for generating the masks.
However, their work focused on temporal compression which is functionally different from the spatial compression task which is the focus of our work.
Inspired by the SDA, a network with an improved architecture was proposed to implement the CS image reconstruction \cite{shi2019image}. Differently from previous reconstruction methods, it is initial reconstruction consisted of multiple $1\times1$ convolutions and a reshape operation. The $1\times1$ convolution, in principle, is functionally equivalent to a fully-connected layer, which fixed the reconstructed image size. After the convolution, the reconstructed 1D vector was reshaped into an initial 2D image. In this work, they experimentally tested their model with binary weights and bipolar weights for image sampling. However, the simple replacement of sampling patterns did not involve the optimization of the overall network. The reported results indicated that the reconstruction accuracy of these two types of weights was sub-optimal compared with their floating-points-based model.

In Section \ref{S:3}, we describe our own network architecture, which aims to solve the aforementioned limitations of the existing methods.

\section{Overview of the proposed network}\label{S:3}
\begin{figure}[ht] \centering \includegraphics[width=\linewidth]{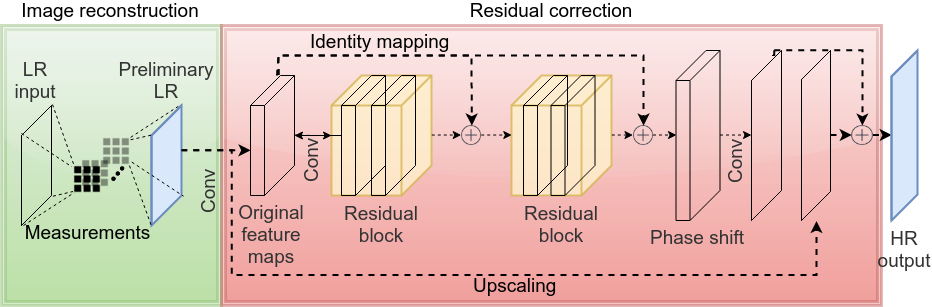}
\caption{The schematic of the proposed network. Our Network has two parts: one that performs image reconstruction and a second part that determines the residual correction. For the image reconstruction part, the network compressively senses the low-resolution input image with static or learned binary patterns and reconstructs the preliminary image. After that, the residual correction sub-net extracts the features from the preliminary image and corrects the reconstruction error using a sequence of recursive residual blocks. Each of these blocks is connected to the original feature maps through identity branches to gradually learn the errors. Then the preliminary image and residuals maps are upscaled through two branches and combined element-wise to generate the final output image.}
\label{fig:network}
\end{figure}

In this section, the network structure is explained in detail. The architecture is shown in Figure \ref{fig:network}. It is functionally divided into two parts, i.e. the \textit{\textbf{image reconstruction sub-net}}, and the \textit{\textbf{residual correction sub-net}}.

Our LSHR scheme assumes an object scene is sampled with low-resolution patterns. In practical applications, ground truth, high-resolution, images are not known a priori. During the training stage, we use the original images as our ground-truth and resample these at low resolution for the purposes of simulating image quality typical of current single pixel imaging systems. These low resolution and ground truth image pairs are used to train our network.

The image reconstruction sub-net samples the low-resolution input images with binary patterns to generate the measurements. From those measurements, the transposed convolution layer learns a non-linear mapping to generate a low-resolution version of the reconstructed image. After that, the residual correction sub-net learns the detail corrections and up-scales the image to the final high-resolution size with a phase shift operation. Together these two parts are able to reconstruct the high-resolution image directly from the low-resolution sampling.

\subsection{Image reconstruction sub-net}
The image reconstruction sub-net learns both the binary patterns and how to reconstruct the image from the measurements. During the training, the sampling process of the computational imaging is done using a convolutional layer where the convolutional kernels act as the digital mirror array and the kernel values (weights) act as binary patterns. When the trained model is integrated with the hardware, the learned kernel values can be uploaded to the digital mirror array to do the sampling and the measurements of the back scattered light intensity are sent back to the network to reconstruct the image.

\begin{figure}[b!] \centering \includegraphics[width=\linewidth]{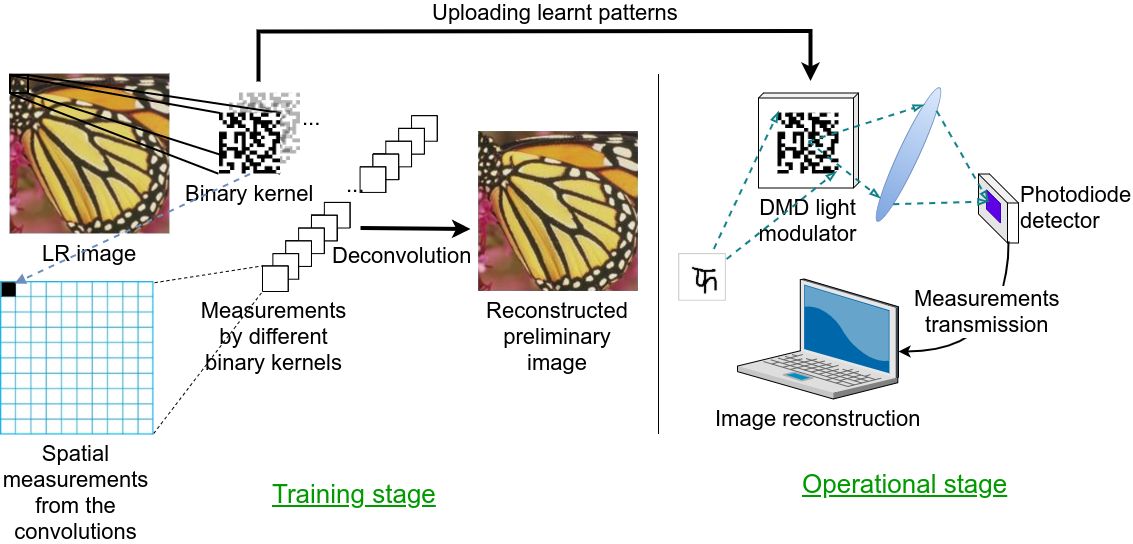}
\caption{The operation of the image reconstruction part. Training stage: the low-resolution image is sampled with binary kernels using a convolutional layer. Each convolution operation generates a measurement, shown as the black element. By sliding the binary kernel through the image, the measurement map for the corresponding binary kernel is generated. After convolution with all binary kernels, the transposed convolutional layer is used to reconstruct the low-resolution preliminary image from the measurements. Operational stage: the learned patterns are uploaded to the DMD hardware to do the sampling, the measurements recorded by the photodiode detector are send back to network to compute the reconstructed image.}
\label{fig:reconstruction}
\end{figure}

The schematic of the image reconstruction sub-net is shown in Figure \ref{fig:reconstruction}. The sampling and reconstruction can be formulated as
\begin{equation}
\begin{aligned}
\tilde{x} 	&=\mathcal{F}_{d}(y, W_{r})+b\\
			&=\mathcal{F}_{d}(\mathcal{F}(\varphi(x),W_{b}), W_{r})+b 
\end{aligned}
\end{equation}
where $\tilde{x}$ is the reconstructed preliminary image. The $\mathcal{F}_{d}(\cdot)$ is the transposed convolution with $W_{r}$ and $b$ are the real-valued kernels and bias respectively. The $\varphi(\cdot)$ down-scales the original images for simulating the sampling process. The measurements $y$ are generated by the convolution $\mathcal{F}(\cdot)$ of image $x$ and the binary kernels $W_{b}$ where each kernel corresponds to a sensing pattern.
In our work, we studied two approaches to generate the binary patterns, i.e. the pre-generated and learned patterns. We describe these in detail below and compare their performance (Section \ref{S:4}).

\textit{Randomly pre-generated binary weights.}
In this approach, the patterns were randomly generated and remained static during the training. Before the training, we initialized the binary weights from the random Bernoulli distribution with $\Pr(1) = 0.5$. The distribution was applied to each kernel independently. During the training process, we updated the weights for the rest of the network. In this approach, the network was trained to fit to a specific set of static binary patterns.
In our experiments, we compared this scheme with the learned binary weights to study the benefit of weight optimization during the training.

\textit{Learned binary weights.}
The kernels were initialized with real-valued weights following the uniform distribution within range $[-1, 1]$. This ensured the initialized weights were equally assigned to positive and negative values. Since the real-valued weights were necessary for the network optimizer during training, these were used for gradient calculation. These were then mapped to binary values and applied to the sensing kernels for forward propagation. The binarization scheme is,
\begin{equation}
w_{b}=\left\{\begin{matrix}
1 \quad if \:\: w_r > 0, \\ 
0 \quad if \:\: w_r \leq 0,
\end{matrix}\right.
\end{equation}
where the $w_{b}$ are the 0, 1 binary weights and the $w_r$ are the real-valued weights. Note that in our network, only the binary kernels were involved in the convolution operations. In addition, we clipped the real-valued weights to fit within the range $[-1, 1]$. This ensured the effective binarizaiton mapping since the very large values out of the range did not have significant impact on the binarization process. We also applied an $\ell_{2}$ norm regularization to the weights to avoid the risk of gradient explosion.

\subsection{Residual correction sub-net} \label{S:3.2}
Taking the output of the image reconstruction sub-net as input, the residual correction sub-net predicts the fine details resulting in a high-resolution output image. The schematic of the residual correction sub-net is shown in the red block in Figure \ref{fig:network}.
This sub-net has two branches: up-scaling and residual mapping. During the training, the upscaling branch interpolates the intermediate input image to the required size of the high-resolution output. The residual mapping branch learns the reconstruction residual (fine details) between the upscaled intermediate input image and the original ground truth image using the long-term recursive residual blocks.
The outputs of the two branches are added element-wise to reconstruct the final high-resolution image.
In the remainder of the section, we describe the long-term recursive residual blocks and the image upscaling processes.

\begin{figure}[t]
\centering \centering \includegraphics[width=0.7\linewidth]{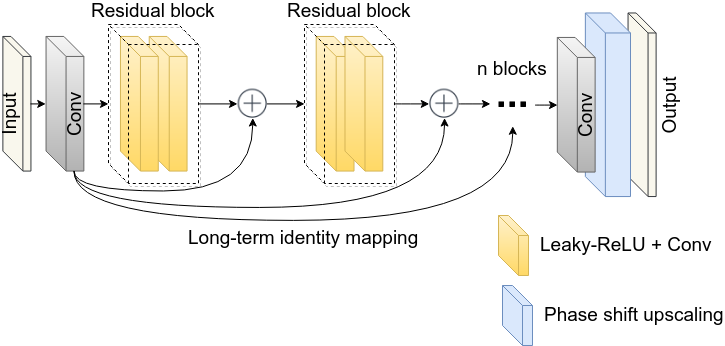}
\caption{The schematic of the residual correction sub-net. The network feeds in the reconstructed preliminary image  as input node and then extracts the original features. The feature maps are then passed to the recursive residual blocks, shown as dashed green lines. Each residual block has an identity branch that connects the original features with its output. Thereafter the residuals and the original features are added, element-wise, to generate the input to the next residual block. For each residual block, we applied leaky ReLU as the pre-activation function. At the end of the network, an extra convolutional layer and an upscaling layer is added to generate the residual output.}
\label{fig:residualnetwork}
\end{figure}

The conventional residual block is formulated as
$\hat{a} = \mathcal{R}(a) = \mathcal{F}(a, W) + h(a)$
where $a$ and $\hat{a}$ are the input and output of the residual block, $W$ indicates the weights of the residual block, $\mathcal{F}(a, W)$ learns the residual mapping between the input and the output and $h(a)$ is the identity mapping function. Our approach differs from the conventional residual block formulation. All of our blocks have skip connections with the intermediate reconstructed images, which we refer to as \textit{long-term connections}. Each block share weights, forming a recursive chain.
The sequence of the blocks in our network is shown in Figure \ref{fig:residualnetwork}. We used two convolutional layers with a pre-activation function in each block.
For the identity mapping, we connected the feature maps associated with the low resolution input (generated by the first convolutional layer) to the output of each block. This long-term connection directly related these features with the outputs of the deep residual blocks. This can be formulated as
\begin{equation}
\begin{aligned}
\hat{a}^j = \mathcal{R}^{j}(\hat{a}^{j-1}) & = \mathcal{F}(\hat{a}^{j-1}, W^j) + h(a^0)\\
\mathcal{F}(\hat{a}^{j-1}, W^j) & = W_{2}^j\sigma(W_{1}^j\sigma(\hat{a}^{j-1}))
\end{aligned}
\end{equation}
where $\mathcal{R}^{j}$ is the residual mapping function of the $j$-th block, $a^0$ is the initial features, and $\hat{a}^j$ is the output of $j$-th block. $W^j$ is the weight and $\sigma$ is the Leaky ReLU activation function \cite{he2015delving}. The $i$th-layer in each block shared the same weights $W_i$ where $i\in{1,2}$. This formed a recursive structure and reduced the total amount of model parameters significantly.

The image upscaling was implemented at the end of the residual correction sub-net. After the residual mapping branch extracted the residual from the preliminary low-resolution image, we applied a phase shift layer \cite{shi2016real} to enlarge the size of the learned residual by a factor of $s$ to have high-resolution residual features. We set the network such that the high-resolution residual features have the same number of channels (one for grayscale and three for RGB) as the final image. In the up-scaling branch, we also enlarged the image size by $s$ with the phase shift operation. Then the residual and the image were added, element-wise, to generate the output image in the high-resolution. In our experiment, we set the upscaling factor $s$ as $2$.

\begin{figure}[t]
\centering \centering \includegraphics[width=\linewidth]{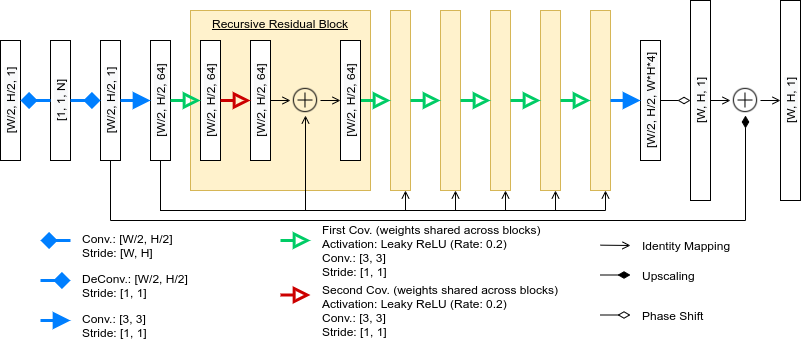}
\caption{The details of the network structure. The diagram illustrates the feature map convolution, in which we take an image of size $H=32$ and $W=32$ as input. The image sampling is done at downscaled resolution and the output is at original resolution. The first yellow block illustrate the inner structure of the recursive residual blocks, which were simplified in the later blocks.}
\label{fig:network_detail}
\end{figure}

\subsection{Network training}
The details of the network structure used in our experiment are illustrated in Figure \ref{fig:network_detail}. The network structure code can be downloaded at our \href{https://github.com/FangliangBai/LSHR-Net}{\textit{GitHub repository}}.
The proposed network consists of two functionally different sub-nets which contain different types of weights respectively. A straightforward strategy, used in previous work, to train such a heterogeneous network, is to train the two parts separately in a pipeline manner. Hence, the image-reconstruction sub-net is first trained and then used as a pre-trained model for training the whole network. This approach can be viewed as either a two-step training strategy or as a semi-decoupled strategy \cite{yao2017dr}. In contrast, we trained the heterogeneous network with pure end-to-end learning. These two parts of the network were trained jointly with a separate learning-rate update scheme for each. Specifically, for the image reconstruction sub-net, we set a larger initial learning rate with faster decay. This encouraged a rapid updating of the binary weights in the early stages of training and a slower update in the later stages, facilitating the residual correction sub-net to recover the fine image details. For the the residual correction sub-net, we initialized a relatively small learning rate with a slower decay rate since the residual correction for the details is more difficult to learn.

Denoting the original image as $x$, we aim to train the whole network $f$ to reconstruct the high-resolution image $\tilde{x}=f(x, W)$, where W denotes the weights of the model. We associated the loss function with the output of both sub-nets (parts), i.e. the reconstructed low-resolution image and the upscaled high-resolution image, to train the network. In contrast to the common $\ell_{2}$-norm loss function, used in previous work, we trained the network using the Charbonnier loss function, which is a variant of the $\ell_{1}$-norm function. Given $\tilde{x}^{s}$ the generated image at $s$ upscaling factor, then our loss function is written as
\begin{equation}
  \pazocal{L}(x,\tilde{x},W)=\frac{1}{N}\sum_{i=1}^{N}\sum_{s=1}^{S}\omega\alpha(x_{i}^{s}-\tilde{x}_{i}^{s}) + \frac{\lambda}{2N} \sum_{W} w^{2}
\end{equation}
where $N$ is the batch size and $\alpha(\mu)=\sqrt{\mu^{2}+\epsilon}$ denotes the Charbonnier penalty. The second term is the $\ell_{2}$-norm regularization for the weights. Our experiments indicated that images generated using the Charbonnier loss function were usually sharper than the results obtained using an $\ell_{2}$ norm loss function. We accumulated the loss of both sub-nets. The ground truth image $x_{i}^{1}$ was generated by downsizing the original image using the bicubic interpolation method. The scalar weight $\omega$ controls the influence of each $x_{i}^{s}$ in the loss function. In our experiment, we set $\omega = 2^{s}$ for each part. This multi-loss function forms a supervision scheme that can control the residual training at each part of the network.

\section{Experiments}\label{S:4}
We conducted a series of tests to study the performance of the network. First, we evaluated the image reconstruction quality  (see Section~\ref{sec:img_rec_res}) on three datasets. 
Our learned and fixed-pattern binary models showed the first and second highest peak signal-to-noise ratio (PSNR) compared to the four methods reviewed in Section~\ref{S:2}. In Section~\ref{sec:model_train_analysis}, we analyze how fixed and learned patterns affected the model training process. Finally, in Section~\ref{sec:rec_effic}, we assess the reconstruction efficiency of the network in comparison with other tested methods.

\subsection{Datasets}\label{S:4_1}
We used the DIV2K image dataset \cite{agustsson2017ntire} for training and validation.
We applied data augmentation to the training images. Specifically, we randomly cropped $50$ small patches of size  $256\times256$ from each of the $800$ images, that comprise the DIV2K dataset, to generate $40,000$ training images. In addition, we randomly applied flipping and rotation to the original patches. We used the cropped image patches as ground truth images for the high-resolution output.

Three datasets were used to test the model's performance.
First we used a benchmark dataset of 11 test images, which has been used in existing work, to evaluate the reconstructed image quality and compare it with the results of previous methods.
Secondly, we evaluated the proposed method on a much larger dataset --  the test set of ILSVRC2017, comprising $50000$ natural images from $1000$ classes \cite{ILSVRC15}.
It is known that natural images are often approximately sparse in the domain of the discrete cosine transformation (DCT) and the wavelet transform \cite{taubman2012jpeg2000}, and CS is an efficient method for approximate recovery of such images.
Since our method is an alternative to CS, we have also evaluated the performance of our structured signal recovery method with images of various levels of sparsity. For this experiment, we generated a DCT-sparse version of the ILSVRC2017 test set and we controlled the sparsity of the DCT coefficients as follows: Each image was first transferred into the DCT domain where the coefficients were reordered based on their magnitude, then we set 5 percentage threshold cases for coefficient magnitude such that $100\%$, $20\%$, $10\%$, $5\%$ or $1\%$ of the coefficients were retained and all other coefficients were set to zero.

\subsection{Setting network parameters and hyperparameters}
\label{sec:hyperparameters}
For the image reconstruction sub-net, we used $16\times16$ patterns for both the sensing kernels and the transposed convolution kernels. For the residual blocks, the kernel size for the convolutional layers was $3\times3$ and we used leaky ReLU activation with leaky rate $p=0.2$. We used $64$ channels for each of the convolutional layers.

The network was trained with a batch size of $16$ using the Adam optimizer for $300$ epochs. For the image reconstruction sub-net, we set the initial learning rate and the decay rate to $1\times10^{-4}$ and $0.25$ respectively. For the residual correction sub-net, we set the initial learning rate and the decay rate to $1\times10^{-5}$ and $0.75$ respectively. We set the decay step to $200,000$. The proposed method was trained on an NVidia GeForce GTX 1080Ti GPU.

In our experiment, we trained the network with different measurement ratios,  $R=\frac{m}{N}$, of $0.01$, $0.10$ and $0.25$, where $m$ is the number of sampling kernels and $N$ is the number of pixels in the sensing images. Accordingly, the number of binary kernels for the 128$\times$128 benchmark sampling images are 164, 1638 and 4096.

\begin{figure*}
\centering 
\begin{subfigure}{0.25\textwidth}
   \centering 
   \includegraphics[height=1.1in]{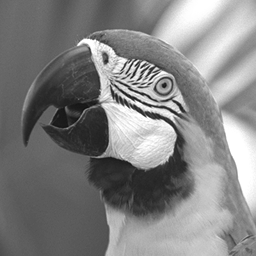}
   \caption{Original \\ \quad} 
   \label{fig:original}
\end{subfigure}%
\begin{subfigure}{0.25\textwidth}
   \centering 
   \includegraphics[height=1.1in]{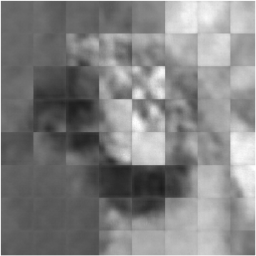}
   \caption{\centering ReconNet (18.93dB)} 
   \label{fig:ReconNet}
\end{subfigure}%
\begin{subfigure}{0.25\textwidth}
   \centering 
   \includegraphics[height=1.1in]{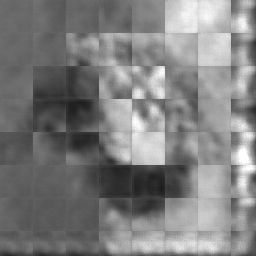}
   \caption{\centering DR$^2$-Net (18.01dB)} 
   \label{fig:DR2net}
\end{subfigure}%
\begin{subfigure}{0.25\textwidth}
   \centering 
   \includegraphics[height=1.1in]{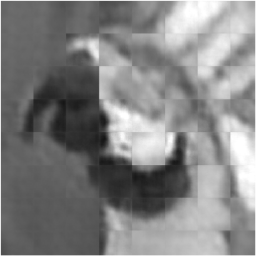}
   \caption{\centering Adp-Rec (21.67dB)} 
   \label{fig:Adp-Rec}
\end{subfigure}%
\\
\begin{subfigure}{0.25\textwidth}
   \centering 
   \includegraphics[height=1.1in]{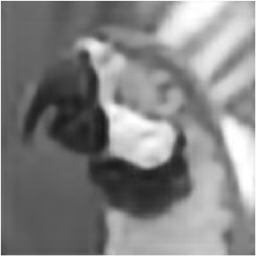}
   \caption{\centering Fully-Conv (22.49dB)} 
   \label{Fully-Rec}
\end{subfigure}%
\begin{subfigure}{0.25\textwidth}
   \centering 
   \includegraphics[height=1.1in]{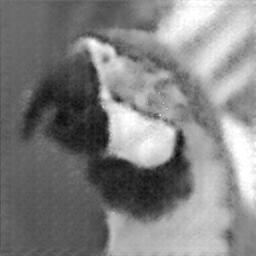}
   \caption{\centering Ours static (22.57dB)} 
   \label{fig:static}
\end{subfigure}%
\begin{subfigure}{0.25\textwidth}
   \centering 
   \includegraphics[height=1.1in]{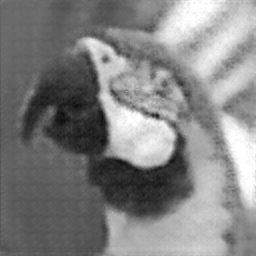}
   \caption{\centering Ours learned (23.01dB)}
   \label{fig:learned}
\end{subfigure}%
   \caption{The reconstruction result of the tested methods, including two of ours, at the compression ratio of $R=0.01$.} 
   \label{fig:result_0.01}
\end{figure*}

\subsection{Image reconstruction results}\label{sec:img_rec_res}
We evaluated our model on the benchmark dataset and compared the results with seven recently proposed methods: ReconNet \cite{kulkarni2016reconnet}, DR$^2$-Net \cite{yao2017dr}, Adp-Rec \cite{xie2017adaptive}, Fully-Conv \cite{xie2017fully}, 2FC2Res \cite{zhao2019visualizing}, Fully-Block Net \cite{xie2018full}, and CSNet$^{+}$ \cite{shi2019image}. To be consistent with previous work, we used the PSNR as the metric.
The comparison results are summarized in Table \ref{tab:snr}. From the table, it can be seen that our network with learned patterns achieved the highest average PSNR at all three measurement ratios. Note the comparison with the Fully-Block Net and CSNet$^{+}$ follows protocols that were reported in their work. Our model with learned patterns indicates better results using the same protocol.

\begin{figure*}
\centering 
\begin{subfigure}{0.25\textwidth}
   \centering 
   \includegraphics[height=1.1in]{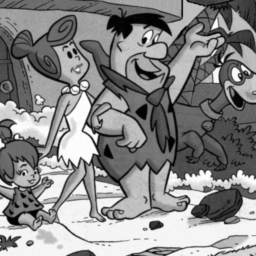}
   \caption{Original \\ \quad} 
\end{subfigure}%
\begin{subfigure}{0.25\textwidth}
   \centering 
   \includegraphics[height=1.1in]{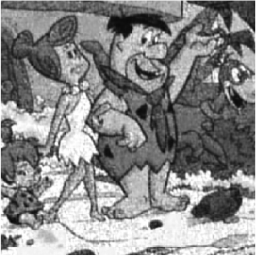}
   \caption{\centering ReconNet (19.04dB)}
\end{subfigure}%
\begin{subfigure}{0.25\textwidth}
   \centering 
   \includegraphics[height=1.1in]{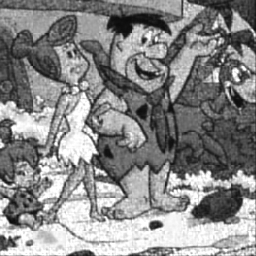}
   \caption{\centering DR$^2$-Net (21.09dB)} 
\end{subfigure}%
\begin{subfigure}{0.25\textwidth}
   \centering 
   \includegraphics[height=1.1in]{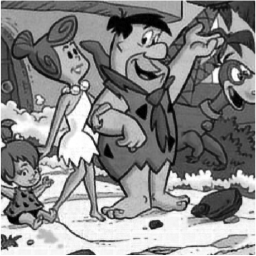}
   \caption{\centering Adp-Rec (23.83dB)} 
\end{subfigure}%
\\
\begin{subfigure}{0.25\textwidth}
   \centering 
   \includegraphics[height=1.1in]{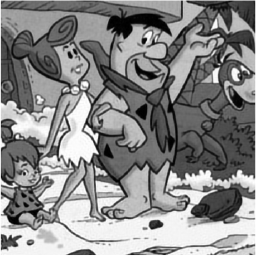}
   \caption{\centering Fully-Conv (24.98dB)} 
\end{subfigure}%
\begin{subfigure}{0.25\textwidth}
   \centering 
   \includegraphics[height=1.1in]{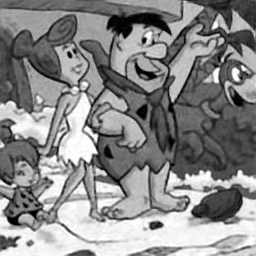}
   \caption{\centering Ours (static) (24.34dB)} 
\end{subfigure}%
\begin{subfigure}{0.25\textwidth}
   \centering 
   \includegraphics[height=1.1in]{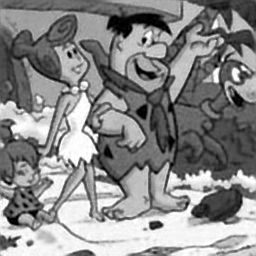}
   \caption{\centering Ours (learned) (24.66dB)}
\end{subfigure}%
   \caption{The reconstruction result of the tested methods, including two of ours, at the compression ratio of $R=0.10$.} 
   \label{fig:result_0.1}
\end{figure*}

\begin{figure*}
\centering 
\begin{subfigure}{0.25\textwidth}
   \centering 
   \includegraphics[height=1.1in]{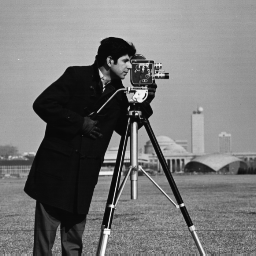}
   \caption{Original \\ \quad} 
\end{subfigure}%
\begin{subfigure}{0.25\textwidth}
   \centering 
   \includegraphics[height=1.1in]{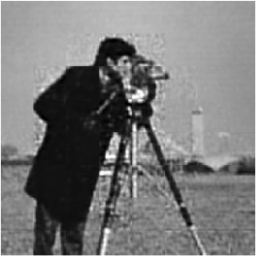}
   \caption{\centering ReconNet (23.48dB)}
\end{subfigure}%
\begin{subfigure}{0.25\textwidth}
   \centering 
   \includegraphics[height=1.1in]{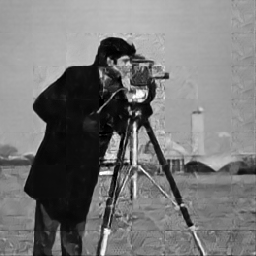}
   \caption{\centering DR$^2$-Net (25.62dB)} 
\end{subfigure}%
\begin{subfigure}{0.25\textwidth}
   \centering 
   \includegraphics[height=1.1in]{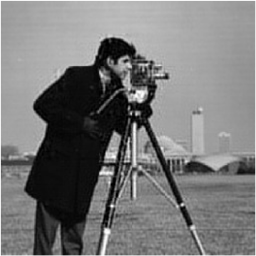}
   \caption{\centering Adp-Rec (27.11dB)} 
\end{subfigure}%
\\
\begin{subfigure}{0.25\textwidth}
   \centering 
   \includegraphics[height=1.1in]{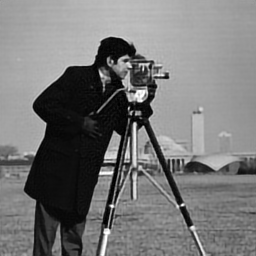}
   \caption{\centering Fully-Conv (28.99dB)} 
\end{subfigure}%
\begin{subfigure}{0.25\textwidth}
   \centering 
   \includegraphics[height=1.1in]{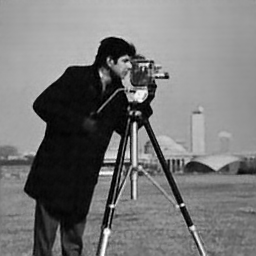}
   \caption{\centering Ours (static) (28.68dB)} 
\end{subfigure}%
\begin{subfigure}{0.25\textwidth}
   \centering 
   \includegraphics[height=1.1in]{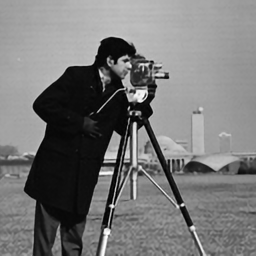}
   \caption{\centering Ours (learned) (30.63dB)} 
\end{subfigure}%
   \caption{The reconstruction result of the tested methods, including two of ours, at the compression ratio of $R=0.25$.} 
   \label{fig:result_0.25}
\end{figure*}

The example images reconstructed by different methods at measurement ratios of $0.01$,$0.10$ and $0.25$ are shown in Figures \ref{fig:result_0.01}, \ref{fig:result_0.1} and \ref{fig:result_0.25} respectively. Our model reconstructed more details than other methods, resulting in images that are visually sharper. At the lowest measurement ratio $0.01$, the block effect is not observed in the output images generated by the Fully-Conv network and our network. This is because both methods used the convolutional layer rather than the fully-connected layer to implement the sensing.
Therefore the network could be trained in an end-to-end fashion and post-processing was not required to smooth the output images. At the measurement ratio of $0.10$, the blocking effect can be eliminated for all methods since a sufficient number of measurements were acquired. At the highest measurement ratio $0.25$, the Fully-Conv network is visually comparable to our method but our learned-weights model still achieved a higher PSNR value.

The difference between the results relating to the static patterns and the learned patterns, of our network, is significant at the measurement ratio of $0.01$. The learned-patterns model achieved better average PSNR and reconstructed more detail. This implies that learning binary weights can help preserve more detail for the same measurement ratio and make the model converge faster, thereby reducing the training time.

\begin{table}[ht]
\centering
\caption{The sample images from reconstruction of the large scale test dataset. The rows denotes the reconstruction at different sparsity in DCT domain. The fist row is the reconstruction of the original images. The second to last rows showing the reconstruction of the sparsity-controlled images. Specifically, the sparsity of the images are at $100\%$, $20\%$, $10\%$, $5\%$ and $1\%$ of the original images. The first column shows the ground truth images and the second to last column show the reconstruction at compression ratio of $0.25$, $0.1$ and $0.01$.}
\label{fig:imagenet}
\resizebox{\columnwidth}{!}{
\begin{tabular}{c|cccc} 
\toprule
\multicolumn{1}{c}{\multirow{2}{*}{Sparsity}} & \multirow{2}{*}{Raw image} & \multicolumn{3}{c}{Reconstruction}  \\ 
\cline{3-5}
\specialrule{0em}{1pt}{1pt}
\multicolumn{1}{c}{}                          &                            & $R=0.25$  & $R=0.1$  & $R=0.01$     \\
\hline\hline
\specialrule{0em}{1pt}{1pt}
original & 
\parbox[c]{1.5in}{\includegraphics[width=1.5in]{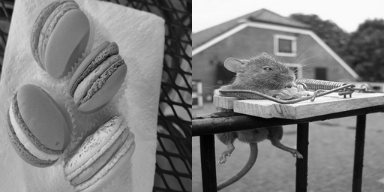}} 	&
\parbox[c]{1.5in}{\includegraphics[width=1.5in]{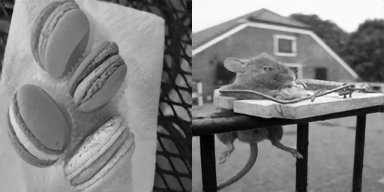}} 	&
\parbox[c]{1.5in}{\includegraphics[width=1.5in]{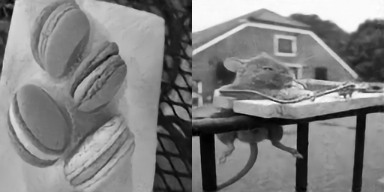}} 	&
\parbox[c]{1.5in}{\includegraphics[width=1.5in]{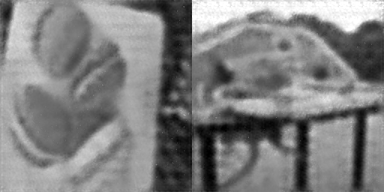}} 	\\
\specialrule{0em}{1pt}{1pt}
\hline
\specialrule{0em}{1pt}{1pt}
20\% &
\parbox[c]{1.5in}{\includegraphics[width=1.5in]{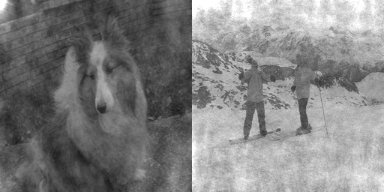}} 	&
\parbox[c]{1.5in}{\includegraphics[width=1.5in]{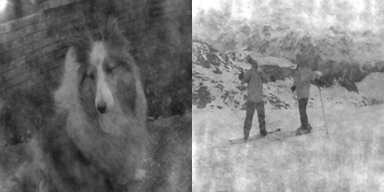}} 	&
\parbox[c]{1.5in}{\includegraphics[width=1.5in]{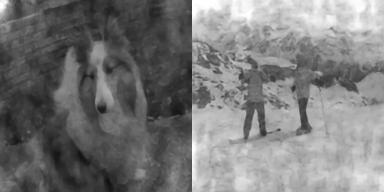}} 	&
\parbox[c]{1.5in}{\includegraphics[width=1.5in]{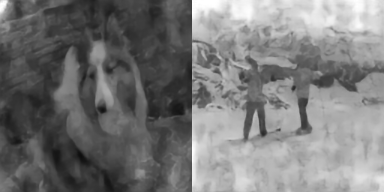}} 	\\
\specialrule{0em}{1pt}{1pt}
\hline
\specialrule{0em}{1pt}{1pt}
10\% &
\parbox[c]{1.5in}{\includegraphics[width=1.5in]{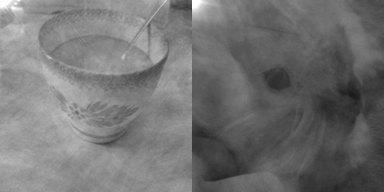}} &
\parbox[c]{1.5in}{\includegraphics[width=1.5in]{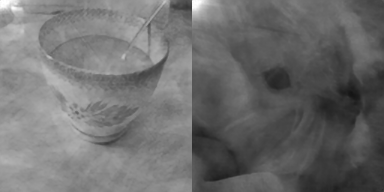}} &
\parbox[c]{1.5in}{\includegraphics[width=1.5in]{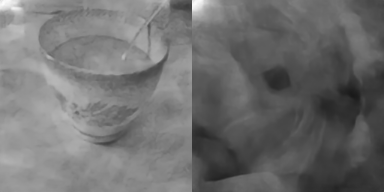}} &
\parbox[c]{1.5in}{\includegraphics[width=1.5in]{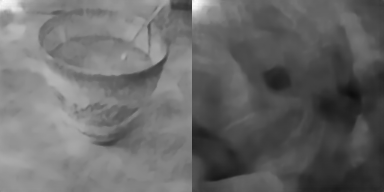}} \\ 
\specialrule{0em}{1pt}{1pt}
\hline
\specialrule{0em}{1pt}{1pt}
5\% &
\parbox[c]{1.5in}{\includegraphics[width=1.5in]{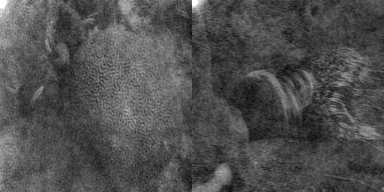}} &
\parbox[c]{1.5in}{\includegraphics[width=1.5in]{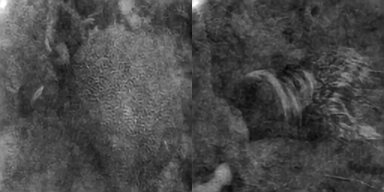}} &
\parbox[c]{1.5in}{\includegraphics[width=1.5in]{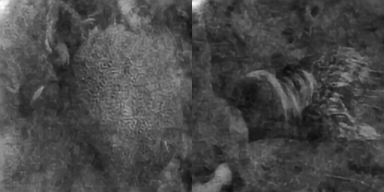}} &
\parbox[c]{1.5in}{\includegraphics[width=1.5in]{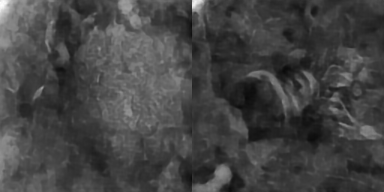}} \\ 
\specialrule{0em}{1pt}{1pt}
\hline
\specialrule{0em}{1pt}{1pt}
1\% & 
\parbox[c]{1.5in}{\includegraphics[width=1.5in]{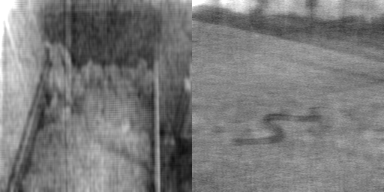}} &
\parbox[c]{1.5in}{\includegraphics[width=1.5in]{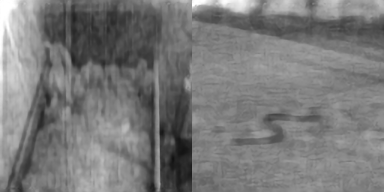}} &
\parbox[c]{1.5in}{\includegraphics[width=1.5in]{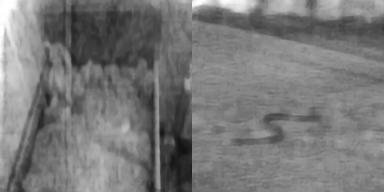}} &
\parbox[c]{1.5in}{\includegraphics[width=1.5in]{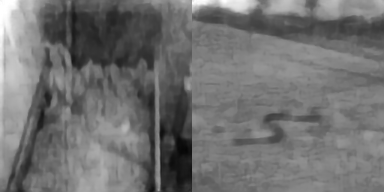}} \\ 
\bottomrule
\end{tabular}
}
\end{table}

\begin{figure}[t] \centering \includegraphics[width=0.7\linewidth]{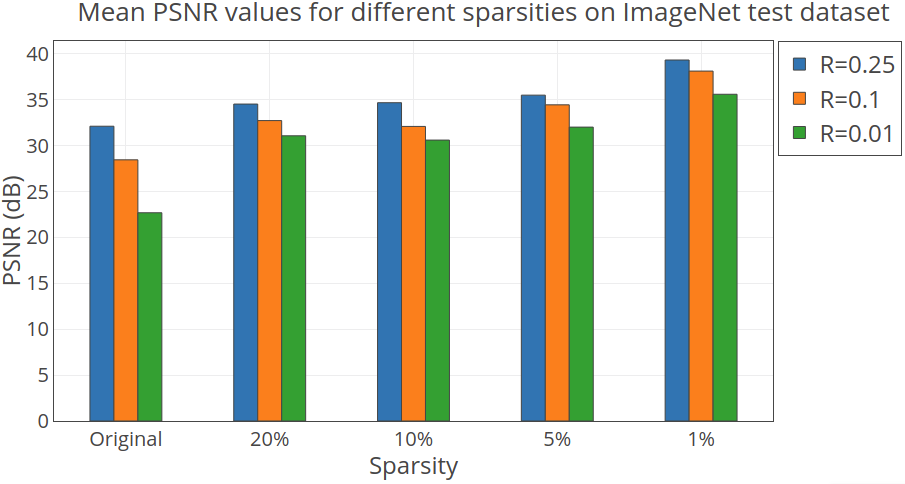}
\caption{The evaluation of the learned binary model on the ILSVRC2017 test set. The trained learned-binary model was tested on the original ILSVRC2017 test set and the sparsified images in three measurement ratios (R = $0.01$, $0.1$ and $0.25$). For the dataset, we controlled the sparsity of the images in the DCT domain. Specifically, we fixed the sparsity of the original images in the DCT domain such that $20\%$, $10\%$, $5\%$ and $1\%$ of the original DCT coefficients were retained. The results show that the trained model works well on the large-scale image dataset, indicating the ability of the model to generalize. It is also observed that the mean PSNR values increase with increasing sparsity. This denotes that the model also performs well on DCT-sparse images.}
\label{fig:sparisity}
\end{figure}

Next, we evaluated the model on the ILSVRC2017 test dataset.
Figure \ref{fig:sparisity} shows the mean PSNR values of the reconstructed images from ILSVRC2017 test set. The mean PSNR values produced by our method in a large-scale test are similar to those produced on a small  benchmark set, indicating good generalization. Furthermore, PSNR values increase with increased sparsity. This indicates that the model performs well also on DCT-sparse images.

We also found that the PSNR of the reconstructed images, at three measurement ratios, tend to be similar when we increase the sparsity of the image in the DCT domain. We present examples of reconstructed images in Table~\ref{fig:imagenet}.

\begin{figure}[t] \centering \includegraphics[width=0.8\linewidth]{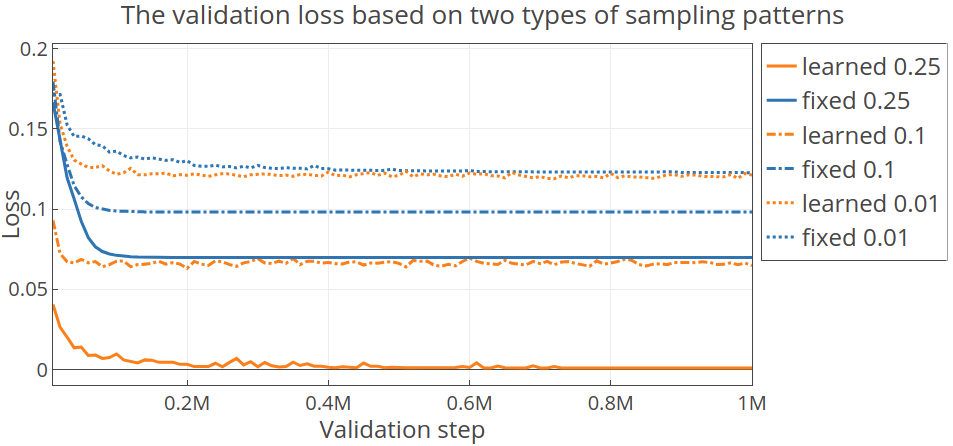}
\caption{The validation losses of models with static (blue lines) and learned (orange lines) binary patterns. Each pattern type was validated for three measurement ratios (R = $0.01$, $0.1$ and $0.25$). 
The validation loss with learned patters drops faster than that with the static patterns. The losses of both models at R = $0.01$ are close at the end of training, but for higher measurement ratio the difference is large.
}
\label{fig:val_loss}
\end{figure}

\subsection{Model training analysis with fixed and binary sampling schemes}\label{sec:model_train_analysis}
First, we analyzed the training efficiency by monitoring the validation loss in both sampling schemes.
We found that training with the learned patterns produced a faster loss reduction for all three measurement ratios (as shown in Figure~\ref{fig:val_loss}) than training with fixed patterns. When the measurement ratio was increased, the discrepancy between the losses of the two networks also increased.
Furthermore, the network with learned patterns yielded a lower final loss, than the fixed patterns network, especially for R of $0.1$ and $0.25$.
Even though the learned patterns network showed some instability compared to the fixed patterns network~\footnote{In the static scheme, the sampling patterns were not involved in the calculation of back-propagation. Only the real-valued weights in the rest of the network were updated. In the learned scheme, the binary weights were updated in each step. The binarization function introduced fluctuations in the gradient calculation, which made the training progress less stable.}, it still is beneficial since it can be trained more quickly. 

\begin{figure}[t] \centering \includegraphics[width=0.8\linewidth]{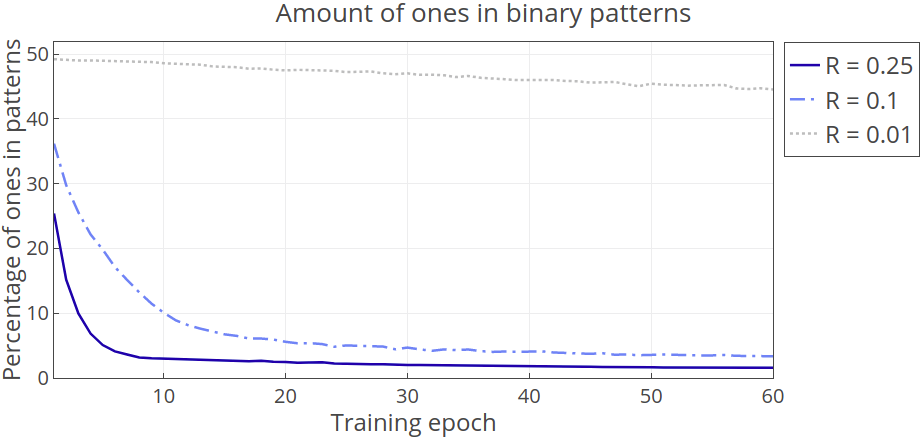}
\caption{During training the binary patterns adapt differently for each measurement ratio. Notice that the fraction of ones contained in the binary patterns is inversely at $R=0.25$, while for the very small measurement ratio $R=0.01$, the fraction of ones remains constant because more information needs to be sensed by each pattern.}
\label{fig:b_ones}
\end{figure}

Next, we analyzed the sparsity of learned patterns by exploring the percentage of valid pixels (with value 1) in the patterns during pattern update.
In compressive sampling theory, we typically use a small number of dense sensing patterns (equal numbers of ones and zeros) in contrast with a raster scan sensing in which each pattern is maximally sparse (contains one on pixel) and records the intensity of single pixel values one at a time. Conversely the sparse patterns are more efficient for single pixel imaging hardware as they require less on-board memory usage. Our approach effectively adapts the sparsity of patterns according to the measurement ratios and hence finds an optimal compromise between sensing efficiency and hardware performance.
Specifically, we initialized all patterns using a single precision uniform distribution within the range $[-1, 1]$ (as required for model optimization), which were subsequently binarized to form patterns with a similar number of ones and zeros.
However, the number of ones decreased dramatically during training since the model at large sampling rates does not necessarily need dense patterns. In contrast, for a relatively small measurement ratio of $R = 0.01$, the number of ones remained consistently high, which suggests that more information was sampled by each pattern.
As a result, the sampling patterns at $R = 0.1$ and $R = 0.25$ contain fewer ones compared to the patterns at $R = 0.01$, as seen in Figure~\ref{fig:b_ones}.
This variation due to $R$ implies that the learning process can generate efficient binary sampling patterns that adapt to different measurements.

\subsection{Analysis of the reconstruction efficiency}\label{sec:rec_effic}
We analyzed the computational efficiency of the network by calculating the time and space complexity, which are introduced in the following content. The results demonstrated that our model has a good balance between the computational cost and the model size for the best image quality.

\begin{table}[t]
\centering
\caption{Efficiency comparison of the tested methods in restoring an image of size 32 $\times$ 32, with the sampling measurement ratio of R=0.01.  The $\mathcal{O}_{space}$ and $\mathcal{O}_{time}$ denote the space and time complexity of the reconstruction layer. The number of convolutional layers and blocks of Fully-Conv were not reported in their work.}
\label{tab:weights_ana}
\resizebox{\columnwidth}{!}{
\begin{tabular}{ccccclcccc}
\toprule
\multicolumn{10}{c}{Reconstruction efficiency and model size of 8 methods}                                                                                 \\ 
\hline
\multicolumn{5}{c}{Image-restoration}                                                     &  & \multicolumn{4}{c}{Residual-correction}                        \\ 
\cline{1-5}\cline{7-10}
Name            & $\mathcal{O}_{space}$  & $\mathcal{O}_{time}$     & \# Weights & Format &  & \# Conv layers & ~Structure & Share weights & Kernel size   \\ 
\hline\hline
ReconNet        & $1.024 \times 10^{4}$  & $1.024 \times 10^{4}$    & 1024       & 32-bit &  & 6              & Plain      & No            & $32\times32$  \\
DR$^2$-Net      & $1.024 \times 10^{4}$  & $1.024 \times 10^{4}$    & 1024       & 32-bit &  & 12             & 4 Blocks   & No            & $32\times32$  \\
Adp-Rec         & $1.024 \times 10^{4}$  & $1.024 \times 10^{4}$    & 1024       & 32-bit &  & 6              & Plain      & No            & $32\times32$  \\
2FC2Res         & $1.024 \times 10^{4}$  & $1.024 \times 10^{4}$    & 1024       & 32-bit &  & 6              & 2 Blocks   & No            & $32\times32$  \\
Fully-Conv      & 2560                   & $2.62144 \times 10^{6}$  & 256        & 32-bit &  & -              & -          & No            & $32\times32$  \\
Fully-Block Net & 2560                   & $2.62144 \times 10^{6}$  & 256        & 32-bit &  & 25             & 12 Blocks  & No            & $32\times32$  \\
CSNet$^{2}$     & $1.024 \times 10^{4}$  & $1.024 \times 10^{4}$    & 1024       & 32-bit &  & 12             & 5 Blocks   & No            & $32\times32$  \\
Ours            & 2560                   & $6.5536 \times 10^{5}$   & 256        & 1-bit  &  & 12             & 6 Blocks   & Yes           & $16\times16$  \\
\bottomrule
\end{tabular}
}
\end{table}

To determine the relative computational efficiency of our network, we compared the model size (space complexity) and the number of operations (time complexity) of our network's image reconstruction layer with the other 4 networks used in prior work (see Table \ref{tab:weights_ana}). The comparison is based on the reconstruction of a single channel (greyscale) image of size $32\times32$ with a measurement ratio $R=0.01$. The comparison is valid for any image size. The time and space complexity are formulated as
$\textbf{Time}\sim \mathcal{O}(M^2\cdot K^2\cdot C_{in}\cdot C_{out})$ and
$\textbf{Space}\sim \mathcal{O}(K^2\cdot C_{in}\cdot C_{out})$, where $M$ is the size of the feature map, $K$ is the size of the kernel, $C_{in}$ and $C_{out}$ are number of input and output channels separately.

Our network has the smallest model size among all the tested networks and lower time complexity than the Fully-Conv network.
Note that the ReconNet, DR$^2$-Net, Adp-Rec, and 2RC2Res perform fewer operations in the initial image reconstruction step because these networks use fully-connected layers. However, the fully-connected layer can only be trained for a specific image size, which is less practical.

For the residual-correction part, our recursive residual block with LSHR sampling scheme generates smaller intermediate feature maps and uses fewer model weights, thereby reducing the computational burden.
In the Fully-Conv and Fully-Block Net networks, images were reconstructed directly back to the high-resolution size. The network then corrected the reconstruction error by applying convolution to the feature maps that had the same size as the high-resolution test image. Since the time complexity is directly related to $M^2$, which is the square of the image size, the computational cost of these three networks increases quadratically when the output image size is doubled.
In contrast, our own network reconstructs the image at low resolution, and then convolutional operations are performed on small feature maps. These are upscaled back to the original size only at the last layer. Therefore, the number of operations performed by our network is order of  $\frac{M^2}{4}$, which is four times less than the Fully-Conv and Fully-Block Net. Furthermore, the number of blocks does not affect the total number of weights since weights are shared between blocks forming a recursive residual block structure. Specifically, the weights are only shared between the first layers (or second layers) among each of the six, two-layer, recursive residual blocks.

\begin{figure}[ht] \centering \includegraphics[width=0.7\linewidth]{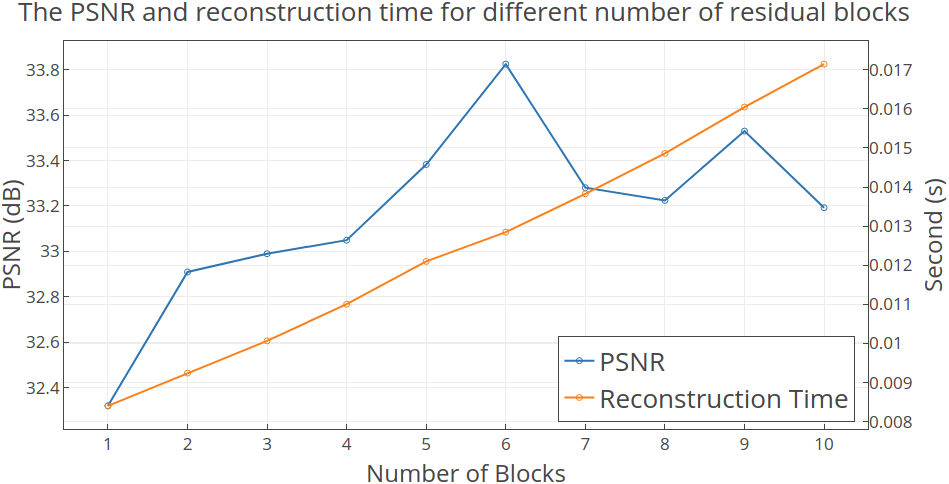}
\caption{The average PSNR  and the reconstruction time as a function of numbers of residual blocks in the recursive structure. The number of residual blocks  influences the performance. The PSNR value was maximized when 6 blocks were used in the recursive structure. The average reconstruction time increased approximately linearly.}
\label{fig:psnr_time}
\end{figure}

The last part of our analysis evaluated the performance of the network for different numbers of residual blocks in our recursive structure.
The depth of the recursive residual block affects the reconstruction accuracy. It is seen in Figure \ref{fig:psnr_time} that the image quality increases by adding more blocks and the best performance (time and accuracy) is obtained with the 6-block structure. Adding more blocks leads to degradation of the image quality. In principle, adding more residual blocks could improve the capability of the residual mapping, but in practice, training a deeper network is harder. 
It is also observed in Figure~\ref{fig:psnr_time} that the reconstruction time increases linearly with the addition of blocks.
Therefore, our final model was constructed by using 6 blocks, which gave the best performance and reasonable reconstruction time.
It was found that the accuracy increased and reached the best performance with 6 blocks, which was used in our final model.

\begin{table}[t]
\centering
\resizebox{\columnwidth}{!}{
\begin{threeparttable}
\caption{The PSNR of 11 test image in dB from recent six methods at three measurement ratios. The reported mean is the average PSNR value for all images. The red figures and the blue figures denote the first and second highest value among all the methods. Our network based on learned binary weights yields the highest average PSNR at all three measurement ratios.}
\label{tab:snr}
\begin{tabular}{cccccccccc} 
\toprule
\multirow{2}{*}{Image}                  & \multirow{2}{*}{Methods} & \multicolumn{3}{c}{\begin{tabular}[c]{@{}c@{}}measurement ratio\\ \end{tabular}}                                    & \multirow{2}{*}{Image}                & \multirow{2}{*}{Methods} & \multicolumn{3}{c}{measurement ratio}                                                                               \\ 
\cline{3-5}\cline{8-10}
                                        &                          & R=0.25                               & R=0.1                                 & R=0.01                               &                                       &                          & R=0.25                               & R=0.1                                & R=0.01                                \\ 
\hline\hline
\multirow{7}{*}{Barbara}                & ReconNet                 & 23.58dB                              & 22.17dB                               & 19.08dB                              & \multirow{7}{*}{Boats}                & ReconNet                 & 27.83dB                              & 24.56dB                              & 18.82dB                               \\
                                        & DR$^2$-Net               & 25.77dB                              & 22.69dB                               & 18.65dB                              &                                       & DR$^2$-Net               & 30.09dB                              & 25.58dB                              & 18.67dB                               \\
                                        & Adp-Rec                  & 27.40dB                              & 24.28dB                               & 21.36dB                              &                                       & Adp-Rec                  & 32.47dB                              & 28.80dB                              & 21.09dB                               \\
                                        & Fully-Conv               & \textcolor[rgb]{0,0.729,1}{28.59dB}  & 24.28dB                               & \textcolor[rgb]{0,0.729,1}{22.06dB}  &                                       & Fully-Conv               & \textcolor[rgb]{0,0.729,1}{33.88dB}  & 29.48dB                              & 22.3dB                                \\
                                        & 2FC2Res                  & 27.92dB                              & 24.27dB                               & 21.48dB                              &                                       & 2FC2Res                  & 33.59dB                              & 29.12dB                              & 21.29dB                               \\
                                        & Ours (static)            & 27.52dB                              & \textcolor{red}{24.57dB}              & 22.03dB                              &                                       & Ours (static)            & 32.05dB                              & \textcolor[rgb]{0,0.729,1}{29.55dB}  & \textcolor[rgb]{0,0.729,1}{22.59dB}   \\
                                        & Ours (Learned)           & \textcolor{red}{31.11dB}             & \textcolor[rgb]{0,0.729,1}{24.56dB }  & \textcolor{red}{22.34dB}             &                                       & Ours (Learned)           & \textcolor{red}{34.13dB}             & \textcolor{red}{29.59dB}             & \textcolor{red}{23.31dB}              \\ 
\hline
\multirow{7}{*}{Fingerprint}            & ReconNet                 & 26.15dB                              & 20.99dB                               & 15.01dB                              & \multirow{7}{*}{Cameraman}            & ReconNet                 & 23.48dB                              & 21.54dB                              & 17.51dB                               \\
                                        & DR$^2$-Net               & 27.65dB                              & 22.03dB                               & 14.73dB                              &                                       & DR$^2$-Net               & 25.62dB                              & 22.46dB                              & 17.08dB                               \\
                                        & Adp-Rec                  & 32.31dB                              & \textcolor[rgb]{0,0.729,1}{26.55dB}   & 16.22dB                              &                                       & Adp-Rec                  & 27.11dB                              & 24.97dB                              & 19.74dB                               \\
                                        & Fully-Conv               & \textcolor[rgb]{0,0.729,1}{32.91dB}  & \textcolor{red}{27.36dB}              & 16.33dB                              &                                       & Fully-Conv               & \textcolor[rgb]{0,0.729,1}{28.99dB}  & 25.62dB                              & 20.63dB                               \\
                                        & 2FC2Res                  & 32.17dB                              & 25.92dB                               & 16.22dB                              &                                       & 2FC2Res                  & 28.84dB                              & 25.07dB                              & 19.98dB                               \\
                                        & Ours (static)            & 30.36dB                              & 26.07dB                               & \textcolor[rgb]{0,0.729,1}{17.10dB}  &                                       & Ours (static)            & 28.68dB                              & \textcolor[rgb]{0,0.729,1}{26.53dB}  & \textcolor[rgb]{0,0.729,1}{20.84dB}   \\
                                        & Ours (Learned)           & \textcolor{red}{33.38dB}             & 26.40dB                               & \textcolor{red}{17.23dB}             &                                       & Ours (Learned)           & \textcolor{red}{30.63dB }            & \textcolor{red}{26.56dB }            & \textcolor{red}{21.35dB}              \\ 
\hline
\multirow{7}{*}{Flinstones}             & ReconNet                 & 22.74dB                              & 19.04dB                               & 14.14dB                              & \multirow{7}{*}{Foreman}              & ReconNet                 & 32.08dB                              & 29.02dB                              & 22.03dB                               \\
                                        & DR$^2$-Net               & 26.19dB                              & 21.09dB                               & 14.01dB                              &                                       & DR$^2$-Net               & 33.53dB                              & 29.20dB                              & 20.59dB                               \\
                                        & Adp-Rec                  & 27.94dB                              & 23.83dB                               & 16.12dB                              &                                       & Adp-Rec                  & 36.18dB                              & \textcolor[rgb]{0,0.729,1}{33.51dB}  & 25.53dB                               \\
                                        & Fully-Conv               & \textcolor[rgb]{0,0.729,1}{30.26dB}  & \textcolor{red}{24.98dB}              & \textcolor[rgb]{0,0.729,1}{16.92dB}  &                                       & Fully-Conv               & \textcolor{red}{38.10dB }            & \textcolor{red}{34.00dB }            & \textcolor{red}{27.26dB}              \\
                                        & 2FC2Res                  & 29.72dB                              & 24.94dB                               & 16.27dB                              &                                       & 2FC2Res                  & 38.25dB                              & 34.29dB                              & 25.77dB                               \\
                                        & Ours (static)            & 28.00dB                              & 24.34dB                               & 16.81dB                              &                                       & Ours (static)            & 35.34dB                              & 33.13dB                              & 26.36dB                               \\
                                        & Ours (Learned)           & \textcolor{red}{31.01dB}             & \textcolor[rgb]{0,0.729,1}{24.66dB}   & \textcolor{red}{17.27dB}             &                                       & Ours (Learned)           & \textcolor[rgb]{0,0.729,1}{36.91dB}  & 33.45dB                              & \textcolor[rgb]{0,0.729,1}{27.13dB}   \\ 
\hline
\multirow{7}{*}{Lena}                   & ReconNet                 & 27.47dB                              & 24.48dB                               & 18.51dB                              & \multirow{7}{*}{House}                & ReconNet                 & 29.96dB                              & 26.74dB                              & 20.30dB                               \\
                                        & DR$^2$-Net               & 29.42dB                              & 25.39dB                               & 17.97dB                              &                                       & DR$^2$-Net               & 31.83dB                              & 27.53dB                              & 19.61dB                               \\
                                        & Adp-Rec                  & 31.63dB                              & 28.50dB                               & 21.49dB                              &                                       & Adp-Rec                  & 34.38dB                              & 31.43dB                              & 22.93dB                               \\
                                        & Fully-Conv               & \textcolor[rgb]{0,0.729,1}{33.00dB}  & 28.97dB                               & 22.51dB                              &                                       & Fully-Conv               & \textcolor[rgb]{0,0.729,1}{36.22dB}  & 32.36dB                              & 23.67dB                               \\
                                        & 2FC2Res                  & 32.97dB                              & 28.86dB                               & 21.57dB                              &                                       & 2FC2Res                  & 35.35dB                              & 31.45dB                              & 22.92dB                               \\
                                        & Ours (static)            & 31.60dB                              & \textcolor[rgb]{0,0.729,1}{29.37dB}   & \textcolor[rgb]{0,0.729,1}{23.13dB}  &                                       & Ours (static)            & 34.80dB                              & \textcolor[rgb]{0,0.729,1}{32.55dB}  & \textcolor[rgb]{0,0.729,1}{24.82dB}   \\
                                        & Ours (Learned)           & \textcolor{red}{34.18dB}             & \textcolor{red}{29.57dB}              & \textcolor{red}{23.52dB}             &                                       & Ours (Learned)           & \textcolor{red}{36.61dB}             & \textcolor{red}{33.73dB}             & \textcolor{red}{25.12dB}              \\ 
\hline
\multirow{7}{*}{Monarch}                & ReconNet                 & 24.95dB                              & 21.49dB                               & 15.61dB                              & \multirow{7}{*}{Peppers}              & ReconNet                 & 25.74dB                              & 22.72dB                              & 17.39dB                               \\
                                        & DR$^2$-Net               & 27.95dB                              & 23.10dB                               & 15.33dB                              &                                       & DR$^2$-Net               & 28.49dB                              & 24.32dB                              & 16.90dB                               \\
                                        & Adp-Rec                  & 29.25dB                              & 26.65dB                               & 17.70dB                              &                                       & Adp-Rec                  & 29.65dB                              & 26.67dB                              & 19.75dB                               \\
                                        & Fully-Conv               & \textcolor[rgb]{0,0.729,1}{32.63dB}  & 27.61dB                               & 18.46dB                              &                                       & Fully-Conv               & 32.90dB                              & \textcolor{red}{28.72dB}             & 21.38dB                               \\
                                        & 2FC2Res                  & 32.46dB                              & 27.60dB                               & 17.85dB                              &                                       & 2FC2Res                  & 32.82dB                              & 27.52dB                              & 20.05dB                               \\
                                        & Ours (static)            & 31.51dB                              & \textcolor[rgb]{0,0.729,1}{28.71dB}   & \textcolor[rgb]{0,0.729,1}{20.09dB}  &                                       & Ours (static)            & \textcolor[rgb]{0,0.729,1}{31.20dB}  & 28.23dB                              & \textcolor[rgb]{0,0.729,1}{21.52dB}   \\
                                        & Ours (Learned)           & \textcolor{red}{34.20dB}             & \textcolor{red}{29.07dB}              & \textcolor{red}{20.79dB}             &                                       & Ours (Learned)           & \textcolor{red}{33.51dB}             & \textcolor[rgb]{0,0.729,1}{28.61dB}  & \textcolor{red}{22.10dB}              \\ 
\hline
\multirow{7}{*}{Parrot}                 & ReconNet                 & 26.66dB                              & 23.36dB                               & 18.93dB                              & \multirow{7}{*}{Mean}                 & ReconNet                 & 26.42dB                              & 23.28dB                              & 17.94dB                               \\
                                        & DR$^2$-Net               & 28.73dB                              & 23.94dB                               & 18.01dB                              &                                       & DR$^2$-Net               & 28.66dB                              & 24.32dB                              & 17.44dB                               \\
                                        & Adp-Rec                  & 30.51dB                              & 27.59dB                               & 21.67dB                              &                                       & Adp-Rec                  & 30.80dB                              & 27.53dB                              & 20.33dB                               \\
                                        & Fully-Conv               & 32.13dB                              & 27.92dB                               & 22.49dB                              &                                       & Fully-Conv               & \textcolor[rgb]{0,0.729,1}{32.69dB}  & 28.30dB                              & 21.27dB                               \\
                                        & 2FC2Res                  & 31.89dB                              & 27.93dB                               & 21.77dB                              &                                       & 2FC2Res                  & 32.36dB                              & 27.91dB                              & 20.47dB                               \\
                                        & Ours (static)            & \textcolor[rgb]{0,0.729,1}{32.64dB}  & \textcolor[rgb]{0,0.729,1}{29.84dB}   & \textcolor[rgb]{0,0.729,1}{22.57dB}  &                                       & Ours (static)            & 31.25dB                              & \textcolor[rgb]{0,0.729,1}{28.44dB}  & \textcolor[rgb]{0,0.729,1}{21.62dB}   \\
                                        & Ours (Learned)           & \textcolor{red}{34.75dB}             & \textcolor{red}{30.18dB}              & \textcolor{red}{23.01dB}             &                                       & Ours (Learned)           & \textcolor{red}{33.68dB}             & \textcolor{red}{28.67dB}             & \textcolor{red}{22.11dB}              \\ 
\hline
\multirow{3}{*}{Mean$^{\diamondsuit}$ } & CSNet \{0,1\}            & -                                    & 26.39dB                               & 20.62dB                              & \multirow{3}{*}{Mean$^{\heartsuit}$ } &                          &                                      &                                      &                                       \\
                                        & CSNet$^{+}$              & -                                    & 28.37dB                               & 21.02dB                              &                                       & Fully-Block Net          & 33.57dB                              & 28.94dB                              & 22.12dB                               \\
                                        & Ours (Learned)           & \textcolor{red}{33.68dB}             & \textcolor{red}{28.67dB}              & \textcolor{red}{22.11dB}             &                                       & Ours (Learned)           & \textcolor{red}{33.66dB}             & \textcolor{red}{29.04dB}             & \textcolor{red}{22.79dB}              \\
\bottomrule
\end{tabular}
\begin{tablenotes}
    \item $\diamondsuit$ Results of CSNet{0,1} and CSNet$^{+}$ at R = 25\% were not reported in their work \cite{shi2019image}.
    \item $\heartsuit$ The Fully-Block Net \cite{xie2018full} was tested only on a subset of the \textit{standard} benchmark. To be specific, seven images from the \textit{standard} benchmark set were selected for testing. To compare with their results, we presented in the table our results on the same subset.
    \end{tablenotes}
\end{threeparttable}
}
\end{table}

\section{Implementation on hardware} \label{S:5}

In real-world applications, the signal/image sampling is usually done by optical devices which inevitably introduce noise and artifacts into the image data. Computer simulations alone provide no guarantees that an image recovery network architecture will be robust to these aspects of practical single-pixel imaging systems. Therefore it is important to validate the efficacy of our LSHR-Net software solution, which uses learned binary patterns, with respect to typical single pixel imaging hardware.

Our hardware comprised a silicon planar photo-detector with a purposely designed amplifier circuit, lenses and a light projector. The photo-detector had a peak sensitivity at the wavelength of $930 nm$ and its sensitive area was $93.6 mm^2$. We connected the circuit to an Arduino circuit board which performed 10-bit analog-to-digital conversion (1024 scales). For evaluation purposes, we used test images from a database as an alternative to setting up unique object scenes. Test images were multiplied, in software, with each of the sampling patterns (forming modulated images) and projected using a TI DLP LightCrafter evaluation module consisting of a built-in DMD plane with a $608\times684$ array. The size, in pixels, of the sampling patterns was constrained by the sensitivity of the photo-detector and the analog-to-digital conversion resolution. A good practical resolution for the sampling patterns was found to be 16x16 pixels. Each of the modulated test images were focused onto the photo-detector using a set of lenses with focal length of $40 mm$, $50 mm$ and $100 mm$. A filter with fixed attenuation was used to reduce light intensity at the photo-detector thereby avoiding saturation. We recorded the light intensity of the modulated images and sent these measurements as inputs data to the model.

For the hardware experiments, we trained our model with the MNIST dataset \cite{lecun-mnisthandwrittendigit-2010} using the same training settings described in Section~\ref{S:4}. The network was trained with $10,000$ MNIST images. The model was evaluated using 18 randomly selected test images of handwritten digits (9 each from MNIST and the Omniglot datasets). We used the Omniglot dataset \cite{lake2015human}, which consists of a set of natural language characters, to demonstrate that the proposed method can generalize to datasets containing images that contains with different image structure from the training set.

\begin{figure}[t]
\centering \includegraphics[width=0.6\linewidth]{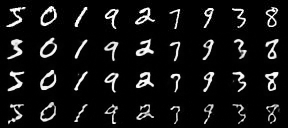}
\caption{The figure shows the reconstruction results of 9 random selected MNIST handwritten numbers using the hardware measurements. The images at the top row are the original ground truth images and the images from the second to the last row are reconstructed results at $R=0.01$, $0.1$ and $0.25$.}
\label{fig:mnist}
\end{figure}

\begin{figure}[t]
\centering \includegraphics[width=0.6\linewidth]{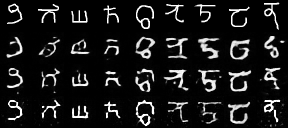}
\caption{The figure shows the reconstruction results of 9 random selected Omniglot characters using the hardware measurements. The images at the top row are the original ground truth images and the images from the second to the last row are reconstructed results at $R=0.01$, $0.1$ and $0.25$.}
\label{fig:omniglot}
\end{figure}

The model reconstructed images directly from the photo-detector measurements at a super resolution size of $32\times32$. We evaluated performance at the same measurement ratios used in Section~\ref{sec:hyperparameters}. Results on MNIST and Omniglot are shown in Figure \ref{fig:mnist} and \ref{fig:omniglot} respectively. It is observed that the reconstruction quality of the character structure was improved by increasing the number of measurements. At the same time, artifacts in the reconstructed images can be seen. These are caused predominantly by noise in the hardware setup (e.g. by the amplifier circuit). The average SNR of the recorded measurement signal was $15.7$dB. Moreover, in Figure \ref{fig:mnist} and \ref{fig:omniglot} it can be seen that the reconstructed images of $R=0.25$ are more pixelated than those of $R=0.1$ and $R=0.01$. Visually, the model resulted in better reconstruction quality. This is however due to the smoothing effect which is also seen in Figure \ref{fig:result_0.01}.

\section{Conclusions}\label{S:6}
In this paper, we have proposed a hardware-friendly method for image reconstruction from compressively sensed measurements, using mixed-weights deep neural networks.
The proposed method, which consists of sampling and reconstruction networks, was specially designed to ease hardware realization, particularly to integrate our work with single pixel camera.
Our novel LSHR network uses trainable binary sampling patterns that can be deployed on a single pixel camera's DMD sampling array.
LSHR net samples light intensity functions at low-resolution and reconstructs images with high-resolution details. Effectively, it reduces the number of measurements at the same measurement ratio and reduces the convolutional computing cost. Hence, it improves the efficiency of the reconstruction process significantly compared with previous work.
For the purpose of reducing the hardware storage requirement for image reconstruction,  the reconstruction network equips long-term recursive residual blocks. It has a weights-sharing strategy that makes the trained models of our method much more compact than those of previously reported network architectures and requires less onboard storage in the imaging hardware.
The experimental results on the benchmark image datasets indicate that our method yields better image quality than those reported in previous work for a number of different measurement ratios.
We also implemented our method on proof-of-concept hardware and demonstrated that it can sample images as compact measurements and then recover them from the measurements successfully.
Our network architecture has potential applications beyond the scope of single pixel imaging. For example, it may be adapted for similar imaging modalities such as coded aperture imaging and structured light sensing. An efficient approach to network training for different imaging modalities may involve transfer learning and this could be the focus of future work in this area. Moreover, for a specific hardware setup, fine-tuning after the initial deployment of hardware can potentially yield improvements in image quality using software alone.

\section*{Acknowledgements}
We dedicate this article to memory of Craig Douglas (Seismicstuff ltd) who designed and made the amplifier circuit used in our hardware experiment. 


\section*{References}
\bibliographystyle{elsarticle-num} 
\bibliography{sample.bib}


\end{document}